# Maastricht and Monetary Cooperation

**CHRIS KIRRANE**

**Abstract**

This paper describes the opportunities and also the difficulties of EMU with regard to international monetary cooperation. Even though the institutional and intellectual assistance to the coordination of monetary policy in the EU will probably be strengthened with the EMU, among the shortcomings of the Maastricht Treaty concerns the relationship between the founder members and those countries who wish to remain outside monetary union. Even though remedies exist, this problem is currently the main threat to monetary cohesion in Europe and also wider economic and political integration. The institutional and intellectual support for monetary cooperation in the G7 is insufficient. The transition to the third stage of EMU will accentuate this particular deficiency

**European and international monetary cooperation**

Monetary issues are not one of the predominant areas of cooperation between the United States and Europe. If a ranking were established, interest rate and exchange rate policies would be secondary to the regulation of financial markets and trade. Nevertheless, cooperation between the United States and Europe and, more generally, between the central banks and the G7 governments has sometimes played a major role, for instance, when the dollar soared in the mid-80s and collapsed in 1994, and during exceptional crises such as in Mexico in 1995.

European monetary unification could end such timid initiatives. The European Central Bank (ECB) will assume the tasks of the central banks of the EU Member States participating in monetary union. The EU Council of Ministers (in consultation with the ECB, European Commission and European Parliament) will make decisions about European involvement in any new global agreement on exchange rates. It is understood that European policymakers and others are concerned about how these institutions will manage the monetary affairs of the newly formed euro area, regardless of the consequences for the cooperation with the rest of the world.

However, many Americans are unaware of this process and this is unfortunate because there is evidence that economic and monetary union (EMU) will actually make international monetary cooperation more necessary. Until the transition to the third stage of the Maastricht process, uncertainties about the timing and composition of monetary union could increase the volatility of exchange rates of European currencies against the dollar. Any tendency for the ECB to liquidate excess dollar reserves inherited from the national central banks could further disrupt the foreign exchange market. The stability of the currency market is a shared priority of both Europe and the United States and it could generate calls for more concerted action and other forms of international monetary cooperation.

The scholarly consensus is to help policy makers agree on a set of adjustments to economic policy. Strictly speaking, policy coordination does not require that policy makers from different countries share the same point of view. Politicians must justify their actions to their constituents and the credibility of their argument could be questioned if they are contradicting each other. So it is preferable that there is broad agreement in order to coordinate monetary policy internationally.

This analysis raises questions about the field of monetary cooperation in Europe and the United States. Even as the institutional and intellectual support for the coordination of monetary policy in Europe will be further strengthened with the third stage of the transition to EMU, this framework will be limited by the relationship between founding member states of the monetary union and those who are not. Although remedies exist, this problem is the main threat to monetary cohesion in Europe and the programme of broader economic and political integration. By comparison, the institutional and intellectual support for transatlantic monetary cooperation and, more generally, monetary cooperation within the G7 remains low. The transition to the third stage of EMU will accentuate these limitations.

**Europe post-EMU**

The advent of the European Central Bank will transform the institutions of monetary cooperation. Article 2 of the Statutes of the European System of Central Banks (composed of the ECB and the central banks of the Member States) gives the primary objective of maintaining price stability (see Kirrane [1993]), or. This formulation shows how the emerging political consensus in the late 80s and 90s facilitated the negotiation of the Maastricht Treaty. It also illustrates the role that the Treaty has played in institutionalising this consensus.

The objective of the ECB will be pursued by a management board that will be appointed as soon as the date of transition to the third stage has been set. It will be composed of six members, including the chair, all appointed by common agreement by the countries participating in monetary union, on the recommendation of the Council of Ministers after consultation with the European Parliament and the Board of Governors of the ECB for a long term of eight years which is nonrenewable.[11] The Board of Governors shall be composed of members of the central bank governors and managers of the participating countries, whose independence has been strengthened and extended by a mandate during the first and second phases. The Board of Governors will take the key decisions on monetary policy, (for example setting interest rates) while the Executive Board will oversee the implementation of these decisions.

To the extent that changes in exchange rates affect the primary objective of the ECB, the Board of Governors will be responsible for foreign exchange policy. At the same time, Article 109 of the Maastricht Treaty empowers the Council of Ministers, acting by qualified majority, to formulate general orientations regarding the exchange policy of non-EU currencies. This provision is likely in part to facilitate the negotiation of agreements of intervention such as the one in the Louvre Agreement. Section 109 provides that these guidelines should not 'affect the objective of maintaining price stability', but does not specify who will decide.

The decision to establish a fixed exchange rate system for industrial countries, as suggested by Paul Volcker (1995), or a global system of 'target-zones', as proposed by John Williamson (1987) will be taken by the Council of Ministers. It will decide unanimously after consulting the ECB and trying to reach a consensus on the compatibility of the decision with price stability. In this case, the Council's decision will be binding on the ECB.

---

[1] Some members will have a shorter term for the initial period.

The inability to adjust national fiscal policies may hamper efforts to coordinate monetary policies. This problem could have relatively serious consequences to the institutional framework of the third stage. While monetary policy of EMU members will be managed by the ECB, fiscal policy will be further defined by the governments of the member states. The excessive deficits procedure (EDP), set out in the Maastricht Treaty, could alleviate this problem. It will be up to the European Council to decide, after a report by the European Commission and the Monetary Committee, if a Member State shows an excessive deficit. The EDP is triggered if the public deficit and public debt represent more than 3% and 60% of GDP 'reference values' as defined in a protocol to the Treaty. Although this procedure is not intended to facilitate the coordination of fiscal policy in itself, it will encourage the exchange of information and analysis of the cross-border impact of national policies. Whether it can be flexibly applied with the judicious coordination of fiscal policies is another matter; if it is rigid, it could actually be a barrier to coordination. [2]

Recognising this problem, the Maastricht Treaty established in section 103, a procedure of mutual surveillance (PMS). It instructs the Council to draft broad economic policies for the Member States, to monitor the development of these policies and make recommendations in case they do not comply with the set guidelines. However, the treaty does not sanction against Member States that do not act as recommended and does not indicate how the PMS and EDP will be coordinated.

The Treaty does not really define the framework for monetary relations between the founding members of EMU and other EU countries. Articles 44 to 47 and 109 provide for the General Council of the ECB, consisting of the Board of Governors and governors of central banks of non-EMU states, but its responsibilities are limited to the collection of statistics and for determining the staff policy. [3]

**Implications for Monetary Cooperation in Europe**

If the Maastricht Treaty only briefly mentions EMS or other coordination mechanisms of monetary policy between EMU members and others, it is because EU officials have not foreseen the current situation. They expected that all EU Member States would be ready to enter the third phase when it started. Otherwise, there was optimism that any stragglers would join the others as soon as possible. Assuming that governments are willing and able to maintain their exchange rates within narrow margins (then 2.25% on either side of the central parity), it was decided that compliance with these margins for two years would be one of the four convergence criteria governing admission to EMU. For all these reasons, it was assumed that the intra-EU exchange rate would be stable.

Today the situation is different. The crisis that hit the EMS in 1992 pushed Italy and the United Kingdom out of the exchange rate mechanism, and the UK wishes to remain outside. It forced two new EU members, Sweden and Finland, which had pegged their currencies to the EMS, to float their currencies. The fluctuation margins were widened from 2.25 to 15%, increasing the potential fluctuation of exchange rates. The debate in Britain and Denmark made it clear that those countries which, under the

---

[2] See Eichengreen and von Hagen (1996). It is now more likely that this problem is with the stability pact which complements the EDP.

[3] Furthermore, the General Council has the right to be informed of the decisions of the Board of Governors. Only the President of the ECB and the governors of the national central banks vote in the General Council. Management Board members cannot vote.

Maastricht Treaty, are subject to an exemption, may decide not to participate in EMU. Successive recessions have shown how difficult it is to meet the criteria on debt and deficit, and focus on the probability of monetary union at different speeds.

For all these reasons, it is now accepted that at the start of EMU, there will be members and non-members. It is therefore necessary to provide mechanisms to promote policy coordination between the two groups.

The PMS is one such mechanism but it does not include countries that do not wish to participate in the monetary union. In the annex to the Treaty there is a requirement that intending members should maintain 'normal' fluctuation margins for two years but it does not define 'normal.' Increasingly, 'normal' is thought to mean 15%. Even this relatively soft constraint is not binding on Member States who do not intend to join EMU immediately.

Hence the spectre of sharp fluctuations in exchange rates between the euro and the currencies of the fringe members of the EU, a prospect that countries like France consider particularly alarming. Paris fears a competitive currency depreciation and the sale of discounted goods from countries like Italy and the UK, who will not be founding members of the monetary union. This is what prompted France to push for a new EMS designed around the single currency. This would be a system of bilateral 'hub and spoke' type relationships in which other currencies are pegged to the euro by bilateral margins, which would be different from the multilateral grid that connects the ERM currencies exchanges today.

The idea that Europe will embark on this path is based on the belief that the political costs of a floating currency is prohibitive, due to the relatively large openness of European economies. The more economies are integrated, the more pronounced will be the effects of internal currency fluctuations. Completing the single market will allow EU countries who devalue their currency to flood the other Member States with their exports. A measure of how well integration progresses, will be how increasingly difficult it is to accept these imports. It will be argued that countries that violate the monetary rules of the Maastricht Treaty cannot enjoy the privileges of the single market. Therefore, uncontrolled fluctuations between the currencies of EMU members and those of other countries could undermine the single market and therefore be unacceptable to all parties concerned. [4]

So powerful is the argument that a system of allowing fluctuations within margins is likely to prevent wild fluctuations of the currencies of countries that do not participate in EMU, the question of the viability of such a system remains. The history of the EMS in the 90s shows how difficult it is to defend exchange rate margins in today's highly liquid markets. Once countries with strong currencies are joined in a monetary union, weak currencies that are not part of the union will be prey to speculators. The ECB will be concerned to establish the credibility of its commitment to price stability so it will hardly be prepared to provide massive support for currencies outside. Monetary union could then be in a fragile financial and monetary environment.

Thus, the prospects for monetary cooperation between the members of EMU and other countries are bleak as the Maastricht Treaty fails to define the institutional framework for cooperation. The crisis that the EMS experienced in 1992-1993 undermined the consensus on economic policy that had prevailed between countries like Sweden and the United Kingdom on the one hand, and France and Germany on the other. These problems have not been resolved.

---

[4] Examples of public statements and others that corroborate this view are given in Eichengreen and Ghironi (1996).

**Implications for transatlantic monetary cooperation**

In the absence of prior planning, EMU threatens to disrupt the institutions of international monetary cooperation. Consultations between the IMF and the EU countries participating in EMU, conducted under Article IV, provide an example. Traditionally, consultations with EU Member States were based on the fact that the review of monetary policies were their responsibility; during the third phase, however, monetary policy will not only be a question individual Member States as fiscal policy will not be under its control, since it is the subject of proceedings concerning excessive deficits and mutual monitoring process of the European Union. These consequences of EMU on the coordination role of the IMF have received minimal attention.[5]

Take the G7 summits. Since 1977, the president of the European Commission participates alongside the leader of the country that chairs the Council of Ministers.[6] In the late 70s, a Board meeting was still held shortly before the summit to try to define a common position and maximise the coherence of the negotiating positions of member countries. This practice has gradually fallen into disuse and coordination between EU and national representatives to the G7 summit has been at best lax.

But neither the president of the European Commission nor the finance minister of the country holding the Presidency of the Council of Ministers will be able to speak on behalf of the European Central Bank which will hold the levers of monetary policy in the euro area. The ECB cannot speak on behalf of Italy and the United Kingdom as the two countries remain outside the monetary union. But it will represent Austria, Ireland and the Benelux countries, which are not part of the G7 (assuming they are founding members of the monetary union).

The President of the Commission will be the designated spokesperson for the EU budgetary authorities if the orientation of fiscal policy of Member States is defined by the implementation of the European Council, the European Commission and the monetary committee concerning excessive deficits and mutual surveillance procedure. Still, the countries with a budget surplus will continue to monitor fiscal policy; in principle, they will also be subject to mutual monitoring process but in practice, it is likely that they will retain considerable budgetary autonomy.

Moreover, there is nothing to suggest that the procedures for excessive deficits and mutual supervision will be strictly enforceable. It must, however, ensure that mutual surveillance procedures also apply during the second phase, even though countries may be reluctant to delegate responsibility for fiscal policy to the Cabinet. Therefore, the overlap of monetary and budgetary powers involves additional complications.

Moreover, attempts to reach agreements such as the Louvre might stumble on the inability of the Council of Ministers and the ECB to agree. Article 109 of the Maastricht Treaty empowers the Council to formulate general orientation regarding foreign exchange policy towards non-Community currencies and includes a clause to allow the negotiation of agreements for intervention in the foreign exchange market. However, Article 109 does not require the ECB to accept and implement the guidelines set by the Council, which it can ignore if they are in conflict with the objective of price stability.[7]

---

[5] Alogoskoufis and Portes (1991) are an exception.

[6] The Commission is involved in discussions on all the issues raised, and not just those directly affecting the European Community. See Putnam and Bayne (1987).

In fact, it is unlikely that monetary cooperation will be the subject of a genuine consensus in the early stages of the third stage. With the creation of an economic and monetary union, the EU will be a large, relatively closed economy like the United States. The bulk of commercial and financial transactions already take place between member states. Increases in transactions within the integrated economic area mean that this will be more so in the future. The effects of fluctuations on exchange rates with the rest of the world will be less troublesome. According to the theory of optimum currency areas, a large and relatively closed economy will be inclined to float its currency. If that is the case, it will be more difficult for the United States to get the ECB and European governments to create a fixed exchange rate system.

At its inception, the ECB will probably hesitate to engage in a policy of concerted intervention in the foreign exchange market as its priority will be to establish its commitment to price stability. Demonstrating an excessive interest in other targets, including the exchange rate, could cast doubt on that commitment. The Board of Governors will tend to strictly interpret Article 109 and to reject the general guidelines of the Council of Ministers if they are considered incompatible with price stability. It is unlikely that monetary policy will be coordinated according to the principles of the Plaza and Louvre agreements during the first years of the third stage. This reality affects fixed exchange rate projects or zonal targets across the G7. The unanimous agreement of the Council of Ministers, necessary before the ECB can participate in such an agreement is a major obstacle.[8] The risk that the Board of Governors raises objections could harm the reputation for financial probity of the Council of Ministers and compromise the viability of the exchange rate agreement and would give the ECB cause to use its veto. The above argument suggests that it would be inclined to exercise this power during the first years of the third stage.

The high probability of a multi-speed EMU further complicates this dynamic. Changes in the dollar exchange rate has long been a problem for EMU. Suppose EMU members and non-members establish a new EMS to limit the movements of foreign currencies against the euro and the European Central Bank takes responsibility for this operation. The latter will then have an interest in stabilising the dollar and could be more favourable to a wider agreement on exchange rate stabilisation. But the limited resources of the ECB and the concerns over the inflationary effects of an intervention would be obstacles to an intervention to limit the decline of the dollar in order to protect the weak ERM currencies. The prospect of such a dilemma could lessen the enthusiasm of the ECB for a broader agreement.

It is conceivable that the advent of the third phase pushes other interest groups to be more concerned about transatlantic monetary cooperation. The French industrial groups, for example, who have always complained of the disruptive effects of currency fluctuations on profitability, will turn their attention away from the mark to the dollar and the yen. This was the case recently, for example, when the franc

---

[7] Kenen (1995, p.32),  it is the ECB that will decide even if this is not in the Treaty

[8] Henning (1996) recommends revising Article 109 to permit a simple majority in the Council

against the mark remained relatively stable, but the fluctuations of the dollar and the yen exacerbated the difficulties of the French economy. In addition, the dollar and the yen will, during the first years of the third phase, be the yardstick by which success the ECB fulfills its mandate in terms of price stability. Whether the euro goes up or down against the dollar will be an obvious measure of success. Although the ECB is reluctant to agree formally to stabilise the euro against the dollar and the yen, it could give sympathetic informal initiatives that would tend towards that goal.

**Strengthen the framework for international monetary cooperation**

Strengthening the framework for international monetary cooperation should be a simple operation in Europe. In comparison with the formidable task of implementing a system of fixed rates or target areas for the G7 as a whole, the experience gained in the context of the EMS and through short-term and very short-term credit facilities should facilitate building a system to stabilise exchange rates between members and non-members of EMU. But a new EMS of the 'hub and spoke' type could weaken the coordination of monetary policies within Europe, not strengthen it. The ECB is unlikely to agree to intervene heavily in the foreign exchange market to support the currencies of non-EMU members who have been kept out of the euro zone because their fiscal and financial policies were questionable.

Presumably, the ECB will estimate that its commitment to price stability is incompatible with support for the financial programmes of governments with dubious reputations. The Maastricht convergence criteria, including the obligation on governments to maintain the value of their currency for two years before entering the third phase, aim at compelling governments to demonstrate financial soundness. The perspectives on concerted market intervention to stabilise exchange rates between the euro and the currencies of non-EMU countries are therefore limited.

According to Charles Wyplosz (1996), the ECB could provide massive support to other Community currencies if non-members agree to submit their economic policies to strict conditionality. Support would only be for countries which are part of the exchange rate mechanism, which respect the obligations of the excessive deficit procedure, which aspire to enter into EMU after a short transition period and are therefore ready to make the necessary adjustments. The viability of such a scheme raises obvious questions. A country wishing to enter EMU could always change its mind. The political means may be lacking when defending a doubling or tripling of interest rates. A country whose currency is attacked could also reconsider its priorities, just as the crisis of 1992 led Sweden to question its participation in the EMS and EMU. All of this will lead the ECB to cover itself. It is unlikely to provide unlimited support. And knowing that the probabilities of such support are slim will only encourage speculators to test the will of policy makers.

The pursuit of an inflation target is the only other option that has focused attention (CEPR 1995). Its supporters claim that interest rates will be stabilised if the ECB adopts a credible operational strategy in this area. The required level of exchange rate stability would be achieved without condemning countries to maintain an unsustainable system of fluctuation margins. This proposal has found some support in the United Kingdom and Sweden, which already pursue an inflation target.[9] But neither of these two countries is part of the 'hard core' of EMU. In key countries such as France and Germany, the intellectual consensus supports aggregate monetary objectives and exchange rates (which suggests, by analogy, that the ECB would give money supply targets to non-member countries in order to define their exchange rates against the euro).

---

[9] See Hamilton *et al* (1996)

In addition, the Maastricht Treaty contains no provision for the adoption of a coordinated inflation target. This could pose some problems of credibility because there is nothing to say that the Member States will be able to meet their inflation targets at the critical time. Without asking questions about what can contribute to this credibility, proponents of inflationary targets tend to assume that such objectives would be credible. They assume that expectations will not be changed by changes in fiscal policy, business cycles or politics.[10] It is fortunate that these questions have not been answered clearly as a concerted inflation objective would provide a better basis for EMS. Unfortunately, the intellectual consensus and the institutional base necessary to put forward such a proposal is lacking.

Possibilities for strengthening transatlantic monetary cooperation are still more limited. In contrast to C. Fred Bergsten and C. Randall Henning (1996), it could be argued that it is unlikely that the G7 countries, or even the USA and Japan could agree on a sustainable system of target areas which would be the centerpiece of a new cooperation within the G7.[11] The idea that Washington could require Article 109 of the Maastricht Treaty to be revised so that unanimity in the Council is no longer a prerequisite for building a new system for the International Monetary Fund is unrealistic. The same applies to the proposals made to revise the complex procedure through which the ECB could agree to participate in a future agreement like the Louvre. It is undeniable that it would be useful for the Council, the ECB and their US counterparts to clarify the conditions in which they would be prepared to consider such operations. The G7 would not be the ideal place for such a process because the President of the European Commission, which participates in its meetings, speaks only indirectly in the name of the Council. Council meetings should therefore again be co-ordinated with G7 summits. This would justify the presence of the President of the Commission and the Minister of Finance of the country holding the presidency of the Council.

At the IMF, bilateral consultations with EU Member States will have to be complemented with consultations with the EU itself, as long as its institutions will determine the monetary policy of EMU members and will influence, through the excessive deficit procedures and the mutual supervision, the budgetary policies of the 15 members. Consultations with EMU member states should not be abolished, however, in the absence of planning these parallel consultations will be a potential source of confusion: for example, the IMF and the EU could give Member States, facing financial and budgetary difficulties, loans with contradictory conditions. It would be as if the IMF consulted not only with the US government but also with each of the 50 US states. It is therefore essential that the IMF and the EU coordinate their action to ensure the compatibility of their conditionality policy.

In principle, EMU should be an opportunity to restructure the representation of the EU at the IMF. Just as a member state of the Federal Reserve System cannot have an unsustainable balance of payments compared to that of another state, the members of EMU will no longer know one another's exchange rate payment problems. It should therefore be the moment to readjust the voting quotas at the IMF. This would reduce the number of votes allocated to each EMU Member State.[12] It would be an opportunity to please Japan, which has long been asking for an increase in its quota and to take into consideration the rise of new commercial powers. The legitimacy of the IMF, which is considered elsewhere in the world as an institution in the service of the interests of the North Atlantic powers, would be increased. For the IMF to retain its legitimacy as one of the major players in the coordination of policies, it is necessary to recognise the changes in international trade flows.[13]

---

[10] Moreover, the followers of this approach do not know the variations of the dollar against the yen, which exceeded 30% in 1994-95, even though the Federal Reserve System and the Bank of Japan pursued policies similar to those of pursuing an inflation objective.

[11] Eichengreen (1994) explains why such a system does not seem feasible.

Some observers believe that international monetary cooperation in future years will focus on the countries facing liquidity crises similar to those which hit Mexico. It may be necessary to mitigate the pressures of a financial crisis by providing assistance as lender of last resort or to lower interest rates. The Mexican episode not only illustrates the desirability of such measures but also barriers to coordination. The reaction of industrialised countries to the Mexican crisis has suffered from the lack of a mechanism which would have collected and deployed the necessary financial resources. This was due to a lack of a sufficient international consensus as to the measures to be taken. The scale of the Mexican crisis required levels of unprecedented funding, so that the United States and other governments of the G7 had to supplement IMF resources on an ad hoc basis (in the case of America, the Clinton Administration used the Stabilisation Fund for Exchange Rates). IMF procedures did not allow the Board of Directors to react with the speed required to prevent a financial collapse. The normal consensus process within the IMF administration had to be bypassed, which displeased some countries and led some European governments to abstain from voting on the loan to Mexico and potentially reduces the prospects for future cooperation.

Some progress has been made on this front. The IMF has established a new emergency funding mechanism to accelerate the disbursement of funds. The G10 and Switzerland have agreed with other creditor countries to double the payments under the General Loan Agreements so that the IMF has the necessary resources. The Gl0 Study Group, composed of central bank governors and finance ministers from major industrialised countries and responsible for analysing the responses to sovereign debt crises took a stand in favour of limited intervention. It recommends modifying loan agreements to allow a restructuring of borrowing conditions by a qualified majority vote and to require the sharing of all payments for the service of the debt.[14] It urged the IMF to consider granting credit even though a country has not paid its arrears. The first recommendations aim to facilitate the orderly restructuring of outstanding debts. The second is intended to provide countries with the necessary working capital to support their banking system and their economy during the process of restructuring. The G10 report also recommends strengthening surveillance at the IMF, accelerating data dissemination and strengthening conditionality of IMF loans.

---

[12] At present, the votes of France, Germany, Austria, Ireland and other countries Benelux are almost similar to those in the United States. Nevertheless, members of EMU could hardly vote in bulk at the IMF since the members of the Executive Board represent groups of countries. Logically, it would therefore be necessary to restructure the groups so that the member countries of EMU form a group in their own right.
[13] One of the problems facing the proponents of a quota review is that EMU is a moving target. The number of participants will continue to change as new members EU will be progressively recognised and that the EU will expand to the East. This reality militates against change.
[14] Group of Ten (1996).

However, much remains to be done.[15] If the G10 governments do not act then clauses relating to representation, sharing payments and qualified majority voting will be added only slowly if at all. Planning a meeting of directors and bondholders will not really facilitate the negotiation; creating standing committees of bondholders remains necessary. If the risks of contagion cannot be proven, the current regulation will continue to apply. As a result, the emergency funds will not be available to a future Mexico, hence the necessity to increase the IMF quota.

**Conclusion**

International monetary cooperation between Europe, the United States and Japan is, at best, sporadic. Institutional mechanisms and the consensus necessary to support more systematic efforts remain insufficient. The advent of the euro and the ECB could be a fatal blow to the relatively limited current arrangements. Some modest measures have been proposed to encourage the creation of institutions and the consensus both inside Europe and on the other side of the Atlantic and the Pacific. However, these recommendations will not mark the beginning a new golden age of international monetary cooperation. Cultivating international monetary cooperation is slow, labourious and progressive process. This is necessarily the case when the preconditions are a more elaborate institutional framework and increased consensus. More extensive cooperation in other areas, such as transatlantic free trade, would help to create comparable links across the Atlantic and thus promote cooperation in monetary matters between the United States and the European Union.

**Bibliography**

Alogoskoufis, George et Richard Portes (1991), International Costs and Benefits from EMU, *European Economy*, numéro spécial, p. 232-245.

Bergsten, C. Fred et C. Randall Henning (1996), *Global Economic Leadership and the Group of Seven*, Institute for International Economics, Washington, DC.

Centre for Economic Policy Research (1995), *Flexible Integration*, CEPR, Londres.

Eichengreen, Barry (1994), *International Monetary Arrangements for the 21st Century*, The Brookings Institution, Washington, DC.

Eichengreen, Barry and Fabio Ghironi (1996), European Monetary Unification: the Challenges Ahead »,in Francisco Torres (ed.), *Monetary Reform in Europe*, Universidade Catholica Editora,Lisbonne, p. 83-120.

Eichengreen, Barry and Richard Portes (1996), *Managing the Next Mexico,* From Halifax to Lyons:*What has Been Done About Crisis Management?*, Peter B. Kenen (ed.), Essays in International Finance, n° 200, International Finance Section, Department of Economics, Princeton University, Princeton, October.

---

[15] The findings are summarised in Eichengreen and Portes (1996).

Eichengreen, Barry and Jûrgen von Hagen (1996), Fiscal Policy and Monetary Union: is There a Tradeoff Between Federalism and Budgetary Restrictions?, *Working Paper*, n° 5517, NBER Working Paper Series, National Bureau of Economic Research, Cambridge, MA, March.

Group of Ten (1996), The Resolution of Sovereign Liquidity Crises, abridged version, *From Halifax to Lyons*: *What has Been Done About Crisis Management?* , Peter B. Kenen (ed.), Essays in International Finance, n° 200, International Finance Section Department of Economics, PrincetownUniversity, Princeton, October.

Hamilton, Carl B., Ulf Jakobsson, lars Jonung, Nils Lundgren and Niels Thygesen (1996),  Swedish Strategies at the European Union Intergovernmental Conference, *SNS Occasional Paper*, March.

Henning, Randall (1994), *Currencies and Politics in the United States, Germany and Japan*, Institute of International Economics, Washington, DC.

Henning, Randall (1996), Europe's Monetary Union and the United States, *Foreign Policy*, n° 102, p.83-102.

Kenen, Peter B. (1995), *Economic and Monetary Union in Europe: Moving Beyond Maastricht*, Cambridge University Press, Cambridge.

Kirrane, Chris. (1993). Lessons from the History of European EMU. *European Economic Integration Review*, 11-22.

Putnam, Robert and Nicholas Bayne (1987), *Hanging Together: Cooperation and Conflict in the Seven Power Summits*, Harvard University Press, Cambridge, Massachussets, revised edition.

Volcker, Paul (1995), The Quest for Exchange Rate Stability: Real or Quixotic?, unpublished manuscript, London School of Economics.

Williamson, John (1987), The Exchange Rate System, *Policy Analysis in International Economics*, n° 5, revised, Institute of International Economics, Washington, DC.

Wyplosz, Charles (1996), An EMS for Both Ins and Outs, unpublished manuscript, Institut universitaire de hautes études internationales, université de Genève.